\documentclass[twocolumn,secnumarabic,amssymb, nobibnotes, aps, prd]{revtex4-2}

\usepackage{graphicx}
\usepackage{dcolumn}
\usepackage{bm}
\usepackage[colorlinks=true,allcolors=blue]{hyperref}
\usepackage{float}
\usepackage{physics}
\usepackage{slashed}
\usepackage{relsize}
\usepackage{amsmath}
\usepackage{hyphenat}

\begin{document}

\title{Pion structure in Holographic QCD}%

\author{Jiali Deng}%
\email{djl2022010355@mails.ccnu.edu.cn}
\affiliation{Institute of Particle Physics and Key Laboratory of Quark and Lepton Physics (MOS),
 Central China Normal University, Wuhan 430079, China}
 \author{Defu Hou}
 \email{houdf@mail.ccnu.edu.cn}
\affiliation{Institute of Particle Physics and Key Laboratory of Quark and Lepton Physics (MOS),
 Central China Normal University, Wuhan 430079, China}
  
 \author{Xiaolong Wang}
\affiliation{Institute of Particle Physics and Key Laboratory of Quark and Lepton Physics (MOS),
 Central China Normal University, Wuhan 430079, China}
 
  \author{Yang Zhou}
\affiliation{Institute of Particle Physics and Key Laboratory of Quark and Lepton Physics (MOS),
 Central China Normal University, Wuhan 430079, China}
  
\date{\today}%

\begin{abstract}
We employ a holographic model with a modified background that incorporates effective descriptions of key QCD features, including linear confinement and gluon condensation, to study the pion's internal structure, encompassing its mass spectrum as well as electromagnetic and gravitational form factors. This model is capable of simultaneously describing these diverse observables and reaches reasonable agreement with both experimental measurements and lattice QCD results. Our findings indicate that the model provides a reasonable description of the pion. The simultaneous reproduction of multiple structure observables supports its potential as a useful tool for further investigations of pion properties.
\end{abstract}
\maketitle




\section{Introduction}\label{sec:01_intro}

As the simplest quark-antiquark bound state and a Nambu-Goldstone boson of spontaneous chiral symmetry breaking, the pion is crucially linked to both strong force and the origin of hadronic mass, playing a pivotal role in QCD (see, e.g., Refs. \cite{Horn:2016rip,Aguilar:2019teb,Ananthanarayan:2022wsl}). Paradoxically, despite being theoretically simpler than the proton, its internal structure remains less well understood \cite{Holt:2010vj}.

A particularly relevant class of models are those based on the anti-de Sitter/conformal field theory (AdS/CFT) correspondence \cite{Maldacena:1997re,Witten:1998qj}. The breaking of conformal invariance yields  effective theories known as holographic QCD or AdS/QCD models. The hard wall model breaks conformal symmetry by introducing a hard cutoff in the holographic coordinate, and the soft wall model is implemented by introducing a soft cutoff via a dilaton field in the action \cite{Polchinski:2001tt,Gherghetta:2009ac,Karch:2006pv}. Alternative methods include modifying the warp factor within the $AdS_5$ metric \cite{FolcoCapossoli:2019imm,Deng:2025fpq,Chen:2025kqb,FolcoCapossoli:2020pks}. These models have successfully addressed many problems in QCD's strongly coupled regime. A key advantage is their ability to incorporate both small and large momentum transfer regimes without an inherent limit on $Q^2$. This stands in contrast to experiments and other approaches like QCD Sum Rules \cite{Tousi:2024usi,Aliev:2023tqy}, which typically are confined to specific momentum transfer windows. This makes the AdS/QCD framework a particularly effective tool for exploring non-perturbative hadron properties beyond the reach of perturbative methods.

These approaches have enabled significant progress on key pion properties. The mass spectrum—a foundational benchmark—has been investigated holographically and by other frameworks, establishing a calibration baseline \cite{deTeramond:2005su,Brodsky:2006uqa,Mondal:2022zlb,Ahmady:2022dfv,Ahmady:2024keo}. Complementing this, the electromagnetic form factors (EMFF) of the pion, which encode its charge distribution, have been investigated via pion electroproduction experiments and through theoretical approaches \cite{Bakulev:2004cu,Brodsky:2007hb,Kwee:2007dd,Chang:2013nia,Brodsky:2014yha,Gao:2021xsm,Chen:2023byr,Puhan:2025pfs,Wang:2025irh}. Similarly, the gravitational form factors (GFFs) of the pion, which reveal its mass and force distributions, remain the subject of active study. Primarily accessed through theoretical tools such as holographic QCD and lattice QCD \cite{Brodsky:2008pf,Xu:2023izo,Polyakov:2018zvc,Broniowski:2024oyk,Cao:2025dkv,Hatta:2025ryj,Choi:2025rto}, they also motivate experimental searches in channels like exclusive meson photoproduction \cite{Seth:2012nn}.

Even with these developments, a fully established holographic model that uses one fixed set of parameters to simultaneously describe the pion mass spectrum, electromagnetic and gravitational form factors—in reasonable agreement with both experiment and lattice data—remains to be fully achieved. This type of integrated analysis provides a way to test the descriptive capacity of a holographic model, namely to see if a single calibrated parameter set can account for various pion characteristics.

In this work, we explore a possible way to approach these problems by modifying the warp factors in the $AdS_5$ metric, rather than introducing a dilaton field in the action. We employ this modified holographic model, which incorporates essential QCD features such as linear confinement and gluon condensation, to investigate the pion's internal structure. We begin by computing the pion's mass spectrum to determine the model parameters. Then, using this set of parameters, we predict the pion's electromagnetic form factors (including the charge radius) and gravitational form factors, and compare them with experimental data and lattice results. The calculated pion structure shows consistency with experimental data and lattice results. Finally, we extend the calculation to explore the dependence of form factors on the pion mass and compute them for pion excited states.

The remainder of the paper is structured as follows: We begin by introducing the soft-wall holographic QCD framework and detailing the pion mass spectrum calculation in Section II. We then predict the pion's electromagnetic form factor in Section III, and derive and analyze the gravitational form factor in Section IV. Next, we extend the calculation to explore the mass dependence of the form factors and compute those for pion excited states in Section V. Finally, Section VI summarizes and discusses the broader
implications of our key findings.

\section{Mass spectrum for pion}\label{sec:02}

In this section, we introduce a phenomenologically corrected holographic model and use it to calculate the mass spectrum of pion mesons. The holographic soft-wall model is the most widely used approach, partly because the soft cutoff it introduces is more natural than the hard cutoff of the hard-wall model, and partly because it can generate linear Regge behavior. In the original soft-wall model, The spacetime metric is given by \cite{Karch:2006pv}
\begin{equation}
	\label{eq1}
ds^2=\frac{1}{z^2}(\eta_{\mu\nu}dx^{\mu}dx^{\nu}+dz^2),
\end{equation}
where $\eta_{\mu\nu}$ denotes the four-dimensional Minkowski metric tensor with signature $(+,-,-,-)$ and $z$ is the holographic coordinate in the fifth dimension, with the boundary at $z=0$ corresponding to the ultraviolet (UV) regime of the dual field theory.
The action can be expressed as:
\begin{equation}
	\label{eq2}
S=\int d^4xdz\sqrt{-g}e^{-\lambda_0(z)}\ell,
\end{equation}
where $\lambda_0(z)=c_0^2z^2$ represents the dilaton field, with $c_0$ a free parameter, and $\ell$ denotes the Lagrangian density. This action can reproduce the confinement properties of QCD. To construct a model that more closely approximates real QCD, an alternative form of the dilaton field can be expressed as \cite{Li:2013oda,Vega:2018dgk,Vega:2020ctz}:
\begin{equation}
	\label{eq3}
\lambda_1(z)=c_1^2z^2tanh(\frac{c_2^4z^2}{c_1^2}),
\end{equation}
where $c_1$ and $c_2$ are free parameters. This model can not only capture the confinement characteristics of QCD, but also simulate gluon condensation.

Although these two models have seen many successful applications \cite{MartinContreras:2018nhv,Brodsky:2008pf,Brodsky:2007hb,Brodsky:2014yha,Vega:2020ctz}, the description of pion meson structure is still incomplete—in particular, these models have difficulty simultaneously describing multiple observables that characterize its internal structure. This suggests that these models may not fully capture the dynamics of pion mesons.

Inspired by Reference \cite{FolcoCapossoli:2020pks}, we model QCD by modifying the $AdS_5$ metric rather than introducing the dilaton field as a prefactor to the action in this study. The key properties of non-perturbative QCD are encoded by this modified background geometry. We assume the form of the modified metric to be
\begin{equation}
\begin{split}
	\label{eq4}
ds^2&=\frac{e^{2k_{1}^{2}z^{2}(1-k_{2}tanh(k_{1}^{2}z^{2}))}}{z^2}(\eta_{\mu\nu}dx^{\mu}dx^{\nu}+dz^2)\\
&=e^{2A(z)}(\eta_{\mu\nu}dx^\mu dx^\nu+dz^2),
\end{split}
\end{equation}
where $k_1$ and $k_2$ are free parameters. The form of warp factor $A(z)$ is
\begin{equation}
	\label{eq5}
A(z)=-\mathrm{ln}(z)+\alpha(z),\ \ \ \alpha(z)=k_{1}^{2}z^{2}(1-k_{2}tanh(k_{1}^{2}z^{2})).
\end{equation}
In this formulation, the $\alpha(z)$ field behaves at the UV boundary as follows:
\begin{equation}
	\label{eq6}
\alpha(z)=k_{1}^{2}z^{2}-k_2k_{1}^{4}z^{4}.
\end{equation}
According to the holographic dictionary, the $z^4$ term maps to the dimension-4 gauge-invariant gluon condensate on the boundary. At the infrared (IR) boundary, it takes the following form:
\begin{equation}
	\label{eq7}
\alpha(z)=k_{1}^{2}z^{2}-k_2k_{1}^{2}z^{2}.
\end{equation}
This indicates that the term which holographically encodes the gluon condensate in the UV evolves into a quadratic form in the IR and thereby contributes to the IR dynamics.

This model incorporates the physical aspects of both forms described above. We first employ this model to calculate the mass spectrum of the pion, from which the model parameters can then be determined.

The five-dimensional action for the pion is \cite{Brodsky:2007hb,Brodsky:2008pf,deTeramond:2013it,Dosch:2025lwb,Brodsky:2014yha}
\begin{equation}
	\label{eq8}
S_\pi=\int d^4xdz\sqrt{-g}[g^{mn}\partial_m\Phi^*\partial_n\Phi-m_5^2\Phi^*\Phi],
\end{equation}
where $\Phi$ represents the field of the pion and $m_5$ is the five dimensional mass of pion. From action (8), one derives the following equations of motion:
\begin{equation}
	\label{eq9}
\partial_m[\sqrt{-g}g^{mn}\partial_n\Phi]-\sqrt{-g}m_5^2\Phi=0,
\end{equation}
where $g^{mn}=e^{-2A(z)}\eta^{mn}$. Eq. (9) can be written as:
\begin{equation}
	\label{eq10}
\partial_m[e^{3A(z)}\eta^{mn}\partial_n\Phi]-e^{5A(z)}m_5^2\Phi=0.
\end{equation}

Defining $\omega(z)=-3A(z)$, one can obtain 
\begin{equation}
	\label{eq11}
\partial_m[e^{-\omega(z)}\eta^{mn}\partial_n\Phi]-e^{\frac{-5\omega(z)}{3}}m_5^2\Phi=0.
\end{equation}

Proceeding, we employ a plane wave ansatz. The amplitude is taken to have dependence only on the holographic coordinate $z$, while propagating in the transverse space $x^\mu$ with momentum $P^\mu$.
\begin{equation}
	\label{eq12}
\Phi(x^\mu,z)=\psi(z) e^{iP_\mu x^\mu},
\end{equation}
where $P^2=P^\mu P_\mu=M_n^2$. A Schrödinger-like equation is obtained after algebraic manipulation and the identification $\psi(z)=e^{\frac{\omega(z)}{2}}\varphi(z)$. The resulting equation is:
\begin{equation}
	\label{eq13}
-\varphi''(z)+[\frac{\omega'(z)^2}{4}-\frac{\omega''(z)}{2}+e^{\frac{-2\omega(z)}{3}}m_5^2]\varphi(z)=M_n^2\varphi(z).
\end{equation}
Here, $M_n$ represents the mass spectrum of the pion in four-dimensional spacetime, where the quantum number $n=1$ corresponds to the ground state, and $n=2,3,...$ denote the excited states.

By solving Equation (13) with the parameter choice $k_1=0.255 GeV, k_2=2.22, m_5^2=-4.095$, we obtain the pion mass spectrum. The 5D pion mass in this study deviates slightly from $m_5^2= -4$ predicted by light-front holographic mapping for a pseudoscalar field with 
$J=L=0$ \cite{Brodsky:2014yha}. The slight deviation naturally reflects the finite quark mass and higher-order contributions, which explicitly break conformal symmetry and effectively shift the 5D mass. Thus, this 5D mass should be understood as an effective mass parameter.

As summarized in Table 1, our calculated mass spectrum for the pion is in reasonable agreement with the experimental data, albeit with some minor deviations. Compared to other studies in the literature, our results show improved accuracy. Consequently, the gravitational background we use provides a closer approximation to real QCD. In the following, we will employ the same set of parameters to compute other observables, thereby further assessing the effectiveness of our model.
\begin{table}
	\centering
	\includegraphics[width=8.5cm]{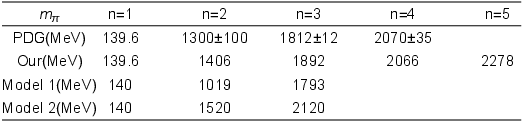}
	\caption{\label{Table 1}Comparison of the pion resonance mass spectrum between experimental values \cite{ParticleDataGroup:2022pth}, other models \cite{Rinaldi:2022dyh,Li:2021jqb}, and our calculations (Our)}
\end{table}

\section{Electromagnetic form factors}\label{sec:04}

As fundamental hadronic observables, form factors encode essential information about hadron structure and dynamics. They capture the critical interplay between perturbative and nonperturbative regimes, a proper description of which should facilitate the study of the transition from high-$Q^2$ (perturbative) to low-$Q^2$ (non-perturbative) dynamics.

In the holographic dual, the electromagnetic form factor is mapped to the coupling between an external EM field $\phi^m(x^\mu,z)$ and a hadron bulk mode $\Phi_P(x^\mu,z)$ in AdS space, as expressed in the following correspondence \cite{Brodsky:2014yha}:
\begin{equation}
\begin{split}
	\label{eq15}
\int &d^4xdz\sqrt{-g}\Phi^*_{P'}(x^\mu,z)\overleftrightarrow{\partial}_m\Phi_{P}(x^\mu,z)\phi^m(x^\mu,z)\\
&\sim (2\pi)^4\delta^4(P'-P-q)\epsilon_\mu(P^\mu+P'^\mu)F(q^2),
\end{split}
\end{equation}
where $P$ and $P'$ are the four-momenta of the initial and final pions, respectively, and $q=P'-P$ is the four-momentum of the virtual photon, with $q^2=-Q^2$. The polarization vector of the photon is denoted by $\epsilon_\mu$, and $F(q^2)$ represents the electromagnetic form factor. Here, $\Phi^* \overleftrightarrow{\partial}_m\Phi=\Phi^*(\partial_m\Phi)-(\partial_m\Phi^*)\Phi$. The lower expression corresponds to the pion EMFF in physical spacetime. It is equivalent to a local coupling with the quark current $J^\mu=\sum e_q \bar{q}\gamma^\mu q$.

The action for the electromagnetic field is
\begin{equation}
	\label{eq16}
S_\gamma=-\frac{1}{4}\int d^4xdz\sqrt{-g}F^{mn}F_{mn},
\end{equation}
where $F_{mn}=\partial_m\phi_n-\partial_n\phi_m$ represents the electromagnetic field strength tensor. The equation of motion corresponding to this action is
\begin{equation}
	\label{eq17}
\partial_m[\sqrt{-g}F^{mn}]=0.
\end{equation}
The gauge condition is
\begin{equation}
	\label{eq18}
\partial_\mu\phi^\mu+e^{-A(z)}\partial_z(e^{A(z)}\phi_z)=0.
\end{equation}

We analyze the propagation of an electromagnetic probe (polarized along the Minkowski coordinates) through AdS spacetime,
\begin{equation}
	\label{eq19}
\phi_\mu(x,z)=\epsilon_\mu e^{iq_\mu x^\mu}V(Q^2,z),\ \ \phi_z=0.
\end{equation}
Its equation of motion is
\begin{equation}
	\label{eq20}
[\partial_z^2+A'(z)\partial_z-Q^2]V(Q^2,z)=0.
\end{equation}
The corresponding boundary conditions are
\begin{equation}
	\label{eq21}
V(Q^2,0)=V(0,z)=1.
\end{equation}
Then, the electromagnetic form factor of the pion can be written as
\begin{equation}
	\label{eq22}
F_\pi(Q^2)=\int dze^{3A(z)}\psi(z)V(Q^2,z)\psi(z).
\end{equation}
\begin{figure}
	\centering
	\includegraphics[width=7cm]{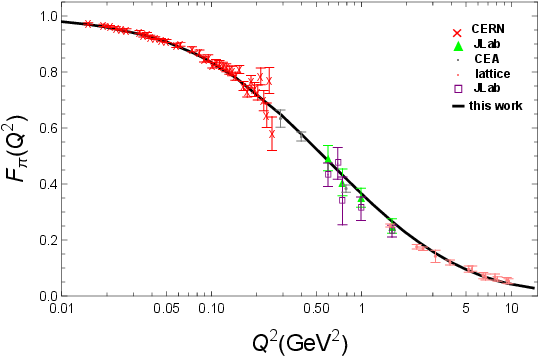}
	\caption{\label{figure}The pion electromagnetic form factor $F_\pi(Q^2)$. The solid black line is our result, the pink points denotes lattice QCD data \cite{Ding:2024lfj}, and the other symbols represent experimental data \cite{NA7:1986vav,Arnold:1974ik,JeffersonLabFpi:2000nlc}.}
\end{figure}

The charge radius, a key observable for probing internal hadron structure, quantifies the spatial extent of the electric charge distribution.

Our analysis is performed in the Breit frame. The relevant kinematic variables—$\Delta^\nu, P^\nu$, and the squared momentum transfer are given by \cite{Dehghan:2025ncw}:
\begin{equation}
	\label{eq23}
P^\nu=(E,\vec{0}),\ \ \Delta^\nu=(0,\vec{\Delta}),\ \ \ \Delta^2=-Q^2,
\end{equation}
where $\Delta^\nu=P'^\nu-P^\nu$. The spatial distribution of the form factor can be obtained via a Fourier transform:
\begin{equation}
	\label{eq24}
F_\pi(r)=\int \frac{d^3\bm{\Delta}}{(2\pi)^3}e^{-i\bm{\Delta}\cdot \textbf{r}}F_\pi(Q^2).
\end{equation}
For the pion, the charge radius is given by:
\begin{equation}
	\label{eq25}
\langle r^2_\pi\rangle=\frac{\int d^3\textbf{r}r^2F_\pi(r)}{\int d^3\textbf{r}F_\pi(r)}=-6\frac{dF_\pi(Q^2)}{dQ^2}|_{Q^2=0}.
\end{equation}
\begin{table}
	\centering
	\includegraphics[width=8.5cm]{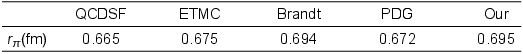}
	\caption{\label{Table 1} The pion charge radius. The first three data points are from lattice QCD \cite{QCDSFUKQCD:2006gmg,Frezzotti:2008dr,Brandt:2013dua}, the fourth is the experimental value \cite{ParticleDataGroup:2010dbb}, and the last one is our result.}
\end{table}

Figure 1 shows the $Q^2$ dependence of the pion electromagnetic form factor. Our calculation is in good agreement with experimental and lattice QCD data. A chi-square test yields $\chi^2/dof\approx 1.12$ for 45 data points, indicating that the deviation between the model and the data is comparable to the experimental  uncertainties. Furthermore, the extracted charge radius of $r_\pi=0.695$ fm is consistent with lattice results and exhibits a mere 3.4\% relative deviation from experiment (see Table 2). These results suggest that our model captures the main electromagnetic properties of the pion to a reasonable extent.

\section{Gravitational form factors}\label{sec:04}

GFFs provide the indispensable framework for exploring hadrons' generalized matter distributions (energy, momentum, pressure) at femtoscale. In the pion—a Nambu-Goldstone boson—these GFFs embed both the basic mass structure generated by dynamical chiral symmetry breaking and the internal mechanical traits that underlie its stability. Thus, computing the pion's GFFs precisely is essential for a first-principles understanding of how mass and spatial structure emerge in QCD. Building on the holographic QCD model constructed previously, this chapter focuses on computing the gravitational form factor A of the pion. Within the holographic duality framework, this form factor corresponds to the coupling between the tensor graviton fluctuation and the scalar field (dual to the pion) in AdS space.

The energy-momentum tensor of the pion can be written as \cite{Brodsky:2008pf}
\begin{equation}
	\label{eq26}
T^{\mu\nu}(x,z)=\frac{-2}{\sqrt{-g}}\frac{\delta S_\pi}{\delta g_{\mu\nu}}.
\end{equation}
Plugging the metric perturbation into the action gives the interaction action that couples the graviton to the pion field.
\begin{equation}
	\label{eq27}
S_{int}=\frac{1}{2}\int d^4xdz\sqrt{-g}h_{\mu\nu}T^{\mu\nu}+ \mathrm{o}(h^2).
\end{equation}
The graviton's equation of motion is obtained from the gravitational action $S_G$ by inserting the metric perturbation $\eta'_{\mu\nu}=\eta_{\mu\nu}+h_{\mu\nu}$. We can get
\begin{equation}
\begin{split}
	\label{eq28}
S_h=\frac{1}{4\kappa^2}\int d^4xdz\sqrt{-g}(\partial_\lambda h^{\mu\nu}\partial^\lambda h_{\mu\nu}-\frac{1}{2}\partial_\lambda h\partial^\lambda h),
\end{split}
\end{equation}
where $h$ represent the trace $h^\mu_\mu$ and $\kappa$ is the Newton constant. By imposing the harmonic-traceless gauge $\partial_\lambda h^{\lambda}_n=\partial_n h=0$, we arrive at the graviton's equation of motion:
\begin{equation}
	\label{eq29}
[\partial_z^2+3A'(z)\partial_z-Q^2]H(Q^2,z)=0,
\end{equation}
where $h_{\mu\nu}=\epsilon_{\mu\nu}e^{-iq_Gx}H(Q^2,z), q_G^2=-Q^2$. The corresponding boundary conditions are:
\begin{equation}
	\label{eq30}
H(Q^2,0)=H(0,z)=1,\ \ \partial_zH(Q^2,\infty)=0.
\end{equation}

The matrix element of the energy-momentum tensor for the pion can be expressed as \cite{Polyakov:2018zvc}
\begin{equation}
	\label{eq31}
\langle P'|T_{\mu\nu}|P\rangle=2P_\mu P_\nu A_\pi(Q^2)+\frac{1}{2}(\Delta_\mu \Delta_\nu-\eta_{\mu\nu}\Delta^2)D_\pi(Q^2),
\end{equation}
where GFFs $A_\pi(Q^2)$ and $D_\pi(Q^2)$ are related to the internal momentum and pressure distributions inside the pion. In this chapter, we restrict our calculation to the GFF $A_\pi(Q^2)$.

We then obtain the gravitational form factor of the pion:
\begin{equation}
	\label{eq32}
A_\pi(Q^2)=\int dze^{3A(z)}\psi(z)H(Q^2,z)\psi(z).
\end{equation}
\begin{figure}
	\centering
	\includegraphics[width=7cm]{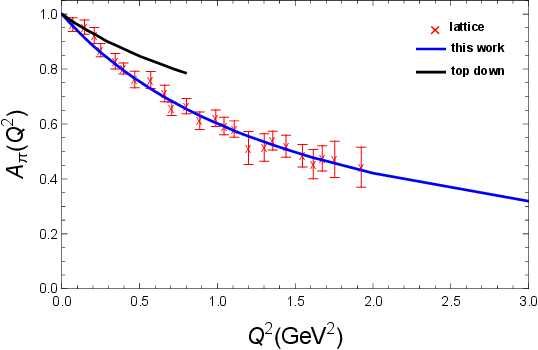}
	\caption{\label{figure}The pion gravitational form factor $A_\pi(Q^2)$ at $m_\pi = 170$~MeV. The
 solid blue line is our result, the solid black line is the top-down result \cite{Fujii:2024rqd}, and the red crosses represent lattice results \cite{Hackett:2023nkr}.}
\end{figure}

As shown in Figure 2, the red crosses denote the lattice QCD results for a pion mass of 170 MeV, while the blue solid line represents our result at the corresponding pion mass. A chi-square test yields $\chi^2/dof\approx 0.39$ for 24 data points, indicating good agreement between the model and the lattice data within uncertainties. From these results, we can extract the radius $r_A \approx 0.39$ fm. This value is close to $r_A = 0.41\pm 0.01$~fm reported from a recent lattice simulation \cite{Hackett:2023nkr}, and also falls within the range $r_A = 0.32\text{--}0.39$~fm obtained from the two-photon process $\gamma^\ast \gamma \rightarrow \pi^0 \pi^0$ \cite{Kumano:2017lhr}.

To assess the robustness of our results against parameter variations, we performed a sensitivity check by varying the two free parameters $k_1$ and $k_2$ individually by $\pm 10\%$ around their optimized values. For the electromagnetic form factor, the average relative deviation from experimental data increases from $2.46\%$ (optimized) to $4.59\%$ ($k_1+10\%$), $4.28\%$ ($k_1-10\%$), $2.93\%$ ($k_2+10\%$), and $2.60\%$ ($k_2-10\%$), respectively. For the gravitational form factor, the deviation from lattice data increases from $2.57\%$ (optimized) to $6.13\%$ ($k_1+10\%$), $7.53\%$ ($k_1-10\%$), $2.78\%$ ($k_2+10\%$), and $3.86\%$ ($k_2-10\%$), respectively. These results demonstrate that the electromagnetic form factor is relatively stable under moderate parameter variations, while the gravitational form factor exhibits somewhat greater sensitivity to $k_1$. Nevertheless, the optimized parameter set consistently yields the best overall description of all observables, confirming the robustness of our holographic framework.

Our model simultaneously describes the pion mass spectrum, electromagnetic form factors, and gravitational form factors, yielding results in agreement with both experimental measurements and lattice QCD calculations. This suggests the potential usefulness of our model. In the next chapter, we will extend our calculation to the form factors of pion excited states and investigate the dependence of the pion form factors on its mass.

\section{EMFF and GFF of pion excited states and their pion-mass dependence}\label{sec:05}

Although the formal expressions for the electromagnetic and gravitational form factors are identical for both ground and excited states—stemming from the same underlying symmetry currents—their physical content differs significantly. While the ground-state pion is well constrained by experiment and serves as a benchmark for model validation, its radial or orbital excitations probe the internal structure of QCD bound states in regimes dominated by higher Fock components, stronger relativistic effects, or altered chiral dynamics. Moreover, investigating the pion’s excited states within our framework not only tests the model’s predictive power beyond the ground state but also provides insights into how hadronic structure evolves with excitation energy and quark mass.

In the following, we present results for the electromagnetic and gravitational form factors of the excited state of the pion, computed using the same formalism as in the ground-state case (see Sections II and III), with the only modification being the replacement of the pion ground-state field by the excited-state field, as obtained from Eq. (13). The corresponding numerical results are shown in Figs. 3 and 4.
\begin{figure}
	\centering
	\includegraphics[width=7cm]{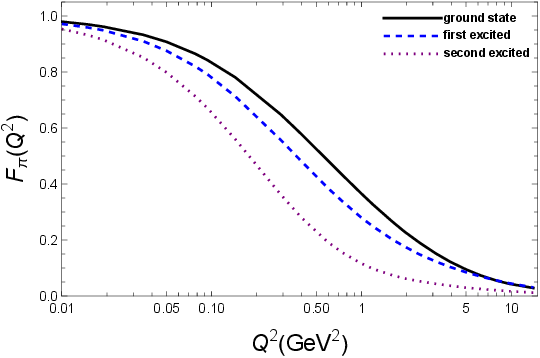}
	\caption{\label{figure}The pion electromagnetic form factor $F_\pi(Q^2)$ for different states. The black line, dashed line and dotted line represent the ground state, the first excited state and the second excited state of pion respectively.}
\end{figure}
\begin{figure}
	\centering
	\includegraphics[width=7cm]{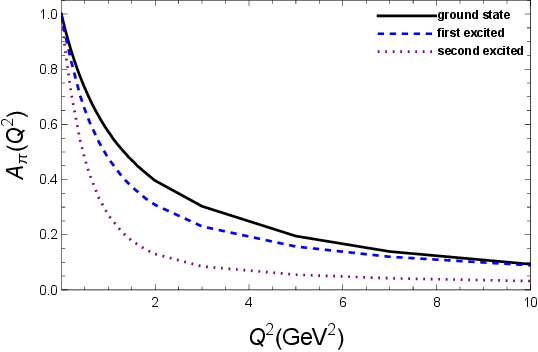}
	\caption{\label{figure}The pion gravitational form factor $A_\pi(Q^2)$ for different states. The black line, dashed line and dotted line represent the ground state, the first excited state and the second excited state of pion respectively.}
\end{figure}

Both the EMFF $F_\pi(Q^2)$ and the GFF $A_\pi(Q^2)$ of the pion satisfy the normalization condition $F_\pi(0)=A_\pi(0)=1$ for ground and excited states alike, reflecting charge conservation and energy-momentum conservation, respectively. Remarkably, the excited-state form factors exhibit a universal qualitative behavior: they fall off more rapidly with increasing momentum transfer $Q^2$ compared to their ground-state counterparts. This means that the excited-state form factor has a larger slope, corresponding to a larger radius, signaling a more spatially extended internal structure—whether in terms of charge or energy-momentum distribution. Despite these differences at low and intermediate $Q^2$, the form factors converge toward the same asymptotic behavior at large $Q^2$. This suggests that, in the deep spacelike regime, the memory of radial excitation is effectively washed out. It is also interesting to note that the excited-state form factors flatten faster at intermediate $Q^2$, indicating a more rapid approach to the asymptotic power-law behavior. These features are clearly illustrated in Figures 3 and 4.

To further investigate the robustness of our model and to explore the dependence of form factors on the pion mass, we extend our analysis to different values of $m_\pi$. Through varying the metric parameter $k_1$, we obtain the ground-state wave functions corresponding to different pion masses by solving Eq. (13). Substituting these wave functions into the formulas then allows us to compute the electromagnetic and gravitational form factors, thereby probing how variations in quark masses affect the internal structure of hadrons. We have calculated the ground-state electromagnetic and gravitational form factors for $m = 0$~MeV, $140$~MeV, and $300$~MeV, respectively, as shown in Figures~5 and~6. This is particularly important for understanding the chiral dynamics within QCD and validating theoretical models across a broader range of parameters. 

\begin{figure}[H]
	\centering
	\includegraphics[width=7cm]{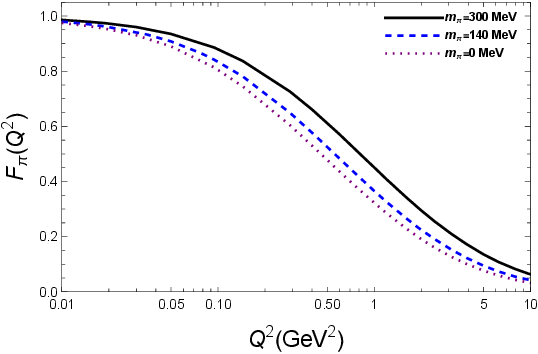}
	\caption{\label{figure}The pion electromagnetic form factor $F_\pi(Q^2)$ for different mass. The black line, dashed line and dotted line represent $m_\pi=300\ \mathrm{MeV}$, $m_\pi=140\ \mathrm{MeV}$ and $m_\pi=0\ \mathrm{MeV}$ respectively.}
\end{figure}
\begin{figure}[H]
	\centering
	\includegraphics[width=7cm]{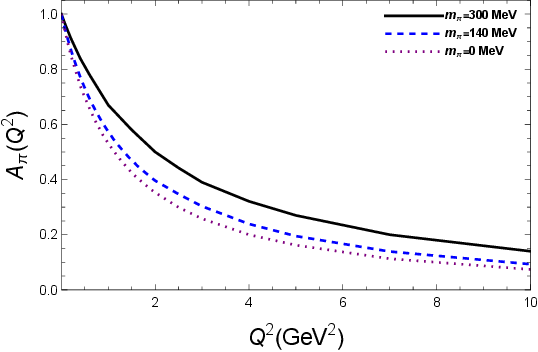}
	\caption{\label{figure}The pion gravitational form factor $A_\pi(Q^2)$ for different mass. The black line, dashed line and dotted line represent $m_\pi=300\ \mathrm{MeV}$, $m_\pi=140\ \mathrm{MeV}$ and $m_\pi=0\ \mathrm{MeV}$ respectively.}
\end{figure}

Both the EMFF $F_\pi(Q^2)$ and the GFF $A_\pi(Q^2)$ of the pion satisfy the normalization condition $F_\pi(0)=A_\pi(0)=1$ for all pion masses considered, as required by charge and energy-momentum conservation. However, at finite momentum transfer $Q^2>0$, we observe a systematic trend: the lighter the pion, the smaller the value of the form factors. This behavior is clearly displayed in Figures 5 and 6. The faster falloff of $F_\pi(Q^2)$ and $A_\pi(Q^2)$ with decreasing pion mass implies a larger slope at $Q^2=0$, which corresponds to a larger root-mean-square radius, thereby signaling a more spatially extended charge and energy-momentum distribution in lighter pions—consistent with the chiral dynamics expectation that the pion becomes increasingly diffuse and extended as $m_\pi\rightarrow 0$. Compared with the lattice results \cite{Koponen:2017fvm,QCDSF:2017ssq}, our extracted charge radius at heavier pion masses is smaller, corresponding to a larger form factor. Despite these quantitative differences, our results reproduce the correct qualitative trend of the lattice data. The observed pion-mass dependence of the form factor suggests that further refinement of the holographic model is needed to fully capture the mass-dependent behavior.

\section{Conclusion}\label{sec:07}

We introduce a phenomenologically deformed holographic model that incorporates key aspects of QCD—such as linear confinement and gluon condensation—and employed it to study the internal structure of the pion. In contrast to the conventional soft wall model, we break conformal invariance by modifying the warp factor of the $AdS_5$ metric. This model includes a term that provides an effective description of the gluon condensate in the high-energy region, which evolves into a quadratic term that also contributes to the infrared dynamics, thus enabling us to better describe the confinement and other QCD properties of the pion.

This work simultaneously calculates the pion mass spectrum, electromagnetic form factors, charge radius, and gravitational form factors, all of which show reasonable agreement with experimental data and lattice QCD results. Subsequently, we extend the study to investigate the form factors of pion excited states and the mass dependence of these form factors. Together, these results suggest that the model provides a viable description.

A limitation of our model is that the modified warp factor is introduced phenomenologically, rather than derived from a dynamical principle. It should be noted that, within the present framework, the decay constants of the pion cannot be directly extracted. The pion decay constant is a fundamental manifestation of chiral symmetry and its breaking, and as such it deserves consideration in the study of pseudoscalar mesons. Its absence from our calculational scheme means that we cannot provide a complete characterization of the pion within our model. In particular, as emphasized in the context of the extended PCAC hypothesis \cite{Dominguez:1976ut}, the decay constants of radially excited states are expected to be strongly suppressed—typically to the few-MeV level—for the excited-state spectrum to be consistent with chiral symmetry breaking. While our framework cannot assess this criterion directly, its potential implications for the reliability of the excited-state properties presented here should be kept in mind. We therefore regard the inability to access decay constants as an inherent limitation of the present phenomenological construction, and we caution the reader that a full validation of our results—especially for excited states—would ultimately require complementary approaches that can determine these quantities. Nevertheless, this phenomenological construction can be justified by its performance: despite its simplified form, it enables a simultaneous description of multiple pion observables—including the mass spectrum, form factors, and charge radius—in agreement with experimental data and lattice QCD results, and can be extended to excited states. The model therefore provides a practical framework for exploring the pion's internal structure.

To conclude, this study indicates that a holographic model incorporating key QCD characteristics can consistently describe a range of pion observables. While some discrepancies remain, the model offers a useful platform for future research on non-perturbative strong interaction physics.

\section*{Acknowledgments}

We thank Hai-cang Ren and  Yan-Qing Zhao for useful discussions. This work is supported in part by the National Key Research and Development Program of China under Contract No. 2022YFA1604900. This work is also partly supported by the National Natural Science Foundation of China(NSFC) under Grants No. 12435009, and No. 12275104.

\bibliographystyle{elsarticle-num}
\bibliography{ref}

@article{Dominguez:1976ut,
    author = "Dominguez, C. A.",
    title = "{Extended Partially Conserved Axial-Vector Current Hypothesis and Chiral Symmetry Breaking}",
    reportNumber = "Print-76-0756 (MEXICO)",
    doi = "10.1103/PhysRevD.15.1350",
    journal = "Phys. Rev. D",
    volume = "15",
    pages = "1350--1360",
    year = "1977"
}

@article{Koponen:2017fvm,
    author = "Koponen, J. and Zimermmane-Santos, A. C. and Davies, C. T. H. and Lepage, G. P. and Lytle, A. T.",
    title = "{Pseudoscalar meson electromagnetic form factor at high $Q^2$ from full lattice QCD}",
    eprint = "1701.04250",
    archivePrefix = "arXiv",
    primaryClass = "hep-lat",
    doi = "10.1103/PhysRevD.96.054501",
    journal = "Phys. Rev. D",
    volume = "96",
    number = "5",
    pages = "054501",
    year = "2017"
}

@article{QCDSF:2017ssq,
    author = "Chambers, A. J. and others",
    collaboration = "QCDSF, UKQCD, CSSM",
    title = "{Electromagnetic form factors at large momenta from lattice QCD}",
    eprint = "1702.01513",
    archivePrefix = "arXiv",
    primaryClass = "hep-lat",
    reportNumber = "ADP-17-06-T1012, DESY-17-019, Edinburgh-2017-02, LTH-1119",
    doi = "10.1103/PhysRevD.96.114509",
    journal = "Phys. Rev. D",
    volume = "96",
    number = "11",
    pages = "114509",
    year = "2017"
}

@article{Fujii:2024rqd,
    author = "Fujii, Daisuke and Iwanaka, Akihiro and Tanaka, Mitsuru",
    title = "{Gravitational form factors of pion from top-down holographic QCD}",
    eprint = "2407.21113",
    archivePrefix = "arXiv",
    primaryClass = "hep-ph",
    doi = "10.1103/PhysRevD.110.L091501",
    journal = "Phys. Rev. D",
    volume = "110",
    number = "9",
    pages = "L091501",
    year = "2024"
}

@article{Kumano:2017lhr,
    author = "Kumano, S. and Song, Qin-Tao and Teryaev, O. V.",
    title = "{Hadron tomography by generalized distribution amplitudes in pion-pair production process $\gamma^* \gamma \rightarrow \pi^0 \pi^0 $ and gravitational form factors for pion}",
    eprint = "1711.08088",
    archivePrefix = "arXiv",
    primaryClass = "hep-ph",
    reportNumber = "J-PARC-TH-0086, KEK-TH-1959, J-PARC-TH-0086 (Erratum: KEK-TH-2096, J-PARC-TH-0155)",
    doi = "10.1103/PhysRevD.97.014020",
    journal = "Phys. Rev. D",
    volume = "97",
    number = "1",
    pages = "014020",
    year = "2018"
}

@article{ParticleDataGroup:2010dbb,
    author = "Nakamura, K. and others",
    collaboration = "Particle Data Group",
    title = "{Review of particle physics}",
    reportNumber = "FERMILAB-PUB-10-665-PPD",
    doi = "10.1088/0954-3899/37/7A/075021",
    journal = "J. Phys. G",
    volume = "37",
    pages = "075021",
    year = "2010"
}

@article{Puhan:2025pfs,
    author = "Puhan, Satyajit and Dahiya, Harleen",
    title = "{Scalar, vector, and tensor form factors of pion and kaon}",
    eprint = "2505.02507",
    archivePrefix = "arXiv",
    primaryClass = "hep-ph",
    doi = "10.1103/2wpb-jgkc",
    journal = "Phys. Rev. D",
    volume = "111",
    number = "11",
    pages = "114039",
    year = "2025"
}

@article{deTeramond:2013it,
    author = {de Teramond, Guy F. and Dosch, Hans G{\"u}nter and Brodsky, Stanley J.},
    title = "{Kinematical and Dynamical Aspects of Higher-Spin Bound-State Equations in Holographic QCD}",
    eprint = "1301.1651",
    archivePrefix = "arXiv",
    primaryClass = "hep-ph",
    reportNumber = "SLAC-PUB-15335",
    doi = "10.1103/PhysRevD.87.075005",
    journal = "Phys. Rev. D",
    volume = "87",
    number = "7",
    pages = "075005",
    year = "2013"
}

@article{Dosch:2025lwb,
    author = "Dosch, Hans Guente and de Teramond, Guy F. and Brodsky, Stanley J.",
    title = "{Holographic light-front QCD}",
    eprint = "2510.20180",
    archivePrefix = "arXiv",
    primaryClass = "hep-ph",
    doi = "10.1016/j.jspc.2026.100339",
    journal = "J. Subatomic Part. Cosmol.",
    volume = "5",
    pages = "100339",
    year = "2026"
}

@article{Vega:2018dgk,
    author = "Vega, Alfredo and Martin Contreras, M. A.",
    title = "{Melting of scalar hadrons in an AdS/QCD model modified by a thermal dilaton}",
    eprint = "1808.09096",
    archivePrefix = "arXiv",
    primaryClass = "hep-ph",
    doi = "10.1016/j.nuclphysb.2019.03.014",
    journal = "Nucl. Phys. B",
    volume = "942",
    pages = "410--418",
    year = "2019"
}

@article{Vega:2020ctz,
    author = "Vega, Alfredo and Martin Contreras, Miguel Angel",
    title = "{Two-body light front wave functions from general AdS/QCD models}",
    eprint = "2005.04501",
    archivePrefix = "arXiv",
    primaryClass = "hep-ph",
    doi = "10.1103/PhysRevD.102.036017",
    journal = "Phys. Rev. D",
    volume = "102",
    number = "3",
    pages = "036017",
    year = "2020"
}

@article{Horn:2016rip,
    author = "Horn, Tanja and Roberts, Craig D.",
    title = "{The pion: an enigma within the Standard Model}",
    eprint = "1602.04016",
    archivePrefix = "arXiv",
    primaryClass = "nucl-th",
    doi = "10.1088/0954-3899/43/7/073001",
    journal = "J. Phys. G",
    volume = "43",
    number = "7",
    pages = "073001",
    year = "2016"
}

@article{Aguilar:2019teb,
    author = "Aguilar, Arlene C. and others",
    title = "{Pion and Kaon Structure at the Electron-Ion Collider}",
    eprint = "1907.08218",
    archivePrefix = "arXiv",
    primaryClass = "nucl-ex",
    reportNumber = "NJU-INP 001/19",
    doi = "10.1140/epja/i2019-12885-0",
    journal = "Eur. Phys. J. A",
    volume = "55",
    number = "10",
    pages = "190",
    year = "2019"
}

@article{Holt:2010vj,
    author = "Holt, Roy J. and Roberts, Craig D.",
    title = "{Distribution Functions of the Nucleon and Pion in the Valence Region}",
    eprint = "1002.4666",
    archivePrefix = "arXiv",
    primaryClass = "nucl-th",
    doi = "10.1103/RevModPhys.82.2991",
    journal = "Rev. Mod. Phys.",
    volume = "82",
    pages = "2991--3044",
    year = "2010"
}

@article{Maldacena:1997re,
    author = "Maldacena, Juan Martin",
    title = "{The Large $N$ limit of superconformal field theories and supergravity}",
    eprint = "hep-th/9711200",
    archivePrefix = "arXiv",
    reportNumber = "HUTP-97-A097, HUTP-98-A097",
    doi = "10.4310/ATMP.1998.v2.n2.a1",
    journal = "Adv. Theor. Math. Phys.",
    volume = "2",
    pages = "231--252",
    year = "1998"
}

@article{Witten:1998qj,
    author = "Witten, Edward",
    title = "{Anti de Sitter space and holography}",
    eprint = "hep-th/9802150",
    archivePrefix = "arXiv",
    reportNumber = "IASSNS-HEP-98-15",
    doi = "10.4310/ATMP.1998.v2.n2.a2",
    journal = "Adv. Theor. Math. Phys.",
    volume = "2",
    pages = "253--291",
    year = "1998"
}

@article{Polchinski:2001tt,
    author = "Polchinski, Joseph and Strassler, Matthew J.",
    title = "{Hard scattering and gauge / string duality}",
    eprint = "hep-th/0109174",
    archivePrefix = "arXiv",
    reportNumber = "NSF-ITP-01-76, UPR-956-T",
    doi = "10.1103/PhysRevLett.88.031601",
    journal = "Phys. Rev. Lett.",
    volume = "88",
    pages = "031601",
    year = "2002"
}

@article{Karch:2006pv,
    author = "Karch, Andreas and Katz, Emanuel and Son, Dam T. and Stephanov, Mikhail A.",
    title = "{Linear confinement and AdS/QCD}",
    eprint = "hep-ph/0602229",
    archivePrefix = "arXiv",
    reportNumber = "BUHEP-06-02, INT-PUB-06-04",
    doi = "10.1103/PhysRevD.74.015005",
    journal = "Phys. Rev. D",
    volume = "74",
    pages = "015005",
    year = "2006"
}

@article{Deng:2025fpq,
    author = "Deng, Jiali and Hou, Defu",
    title = "{Nucleon structure from an AdS/QCD model in the Veneziano limit}",
    eprint = "2502.00771",
    archivePrefix = "arXiv",
    primaryClass = "nucl-th",
    doi = "10.1103/b9p6-pdwh",
    journal = "Phys. Rev. D",
    volume = "112",
    number = "3",
    pages = "036011",
    year = "2025"
}

@article{FolcoCapossoli:2019imm,
    author = "Folco Capossoli, Eduardo and Mart{\'\i}n Contreras, Miguel Angel and Li, Danning and Vega, Alfredo and Boschi-Filho, Henrique",
    title = "{Hadronic spectra from deformed AdS backgrounds}",
    eprint = "1903.06269",
    archivePrefix = "arXiv",
    primaryClass = "hep-ph",
    doi = "10.1088/1674-1137/44/6/064104",
    journal = "Chin. Phys. C",
    volume = "44",
    number = "6",
    pages = "064104",
    year = "2020"
}

@article{FolcoCapossoli:2020pks,
    author = "Folco Capossoli, Eduardo and Mart{\'\i}n Contreras, Miguel Angel and Li, Danning and Vega, Alfredo and Boschi-Filho, Henrique",
    title = "{Proton structure functions from an AdS/QCD model with a deformed background}",
    eprint = "2007.09283",
    archivePrefix = "arXiv",
    primaryClass = "hep-ph",
    doi = "10.1103/PhysRevD.102.086004",
    journal = "Phys. Rev. D",
    volume = "102",
    number = "8",
    pages = "086004",
    year = "2020"
}

@article{Tousi:2024usi,
    author = "Tousi, M. Shekari and Azizi, K. and Moshfegh, H. R.",
    title = "{Investigation of the semileptonic decay {\ensuremath{\Xi}}cc++{\textrightarrow}{\ensuremath{\Xi}}c+{\ensuremath{\ell}}{\textasciimacron}{\ensuremath{\nu}}{\ensuremath{\ell}} within QCD sum rules}",
    eprint = "2409.00241",
    archivePrefix = "arXiv",
    primaryClass = "hep-ph",
    doi = "10.1103/PhysRevD.110.114001",
    journal = "Phys. Rev. D",
    volume = "110",
    number = "11",
    pages = "114001",
    year = "2024"
}

@article{Polyakov:2018zvc,
    author = "Polyakov, Maxim V. and Schweitzer, Peter",
    title = "{Forces inside hadrons: pressure, surface tension, mechanical radius, and all that}",
    eprint = "1805.06596",
    archivePrefix = "arXiv",
    primaryClass = "hep-ph",
    doi = "10.1142/S0217751X18300259",
    journal = "Int. J. Mod. Phys. A",
    volume = "33",
    number = "26",
    pages = "1830025",
    year = "2018"
}

@article{Mondal:2022zlb,
    author = "Mondal, Chandan",
    title = "{Longitudinal dynamics in light-front holographic QCD and hadron spectroscopy}",
    doi = "10.31349/SuplRevMexFis.3.0308091",
    journal = "Rev. Mex. Fis. Suppl.",
    volume = "3",
    number = "3",
    pages = "0308091",
    year = "2022"
}

@article{Ahmady:2022dfv,
    author = "Ahmady, Mohammad and Kaur, Satvir and Mondal, Chandan and Sandapen, Ruben",
    title = {{Pion spectroscopy and dynamics using the holographic light-front Schr{\"o}dinger equation and the 't Hooft equation}},
    eprint = "2208.08405",
    archivePrefix = "arXiv",
    primaryClass = "hep-ph",
    doi = "10.1016/j.physletb.2022.137628",
    journal = "Phys. Lett. B",
    volume = "836",
    pages = "137628",
    year = "2023"
}

@article{Ahmady:2024keo,
    author = "Ahmady, Mohammad and Sandapen, Ruben",
    title = "{Predictions for pion and its excited states using light-front QCD holography and 't Hooft equation}",
    doi = "10.22323/1.476.0540",
    journal = "PoS",
    volume = "ICHEP2024",
    pages = "540",
    year = "2025"
}

@article{Brodsky:2014yha,
    author = "Brodsky, Stanley J. and de Teramond, Guy F. and Dosch, Hans Gunter and Erlich, Joshua",
    title = "{Light-Front Holographic QCD and Emerging Confinement}",
    eprint = "1407.8131",
    archivePrefix = "arXiv",
    primaryClass = "hep-ph",
    reportNumber = "SLAC-PUB-15972",
    doi = "10.1016/j.physrep.2015.05.001",
    journal = "Phys. Rept.",
    volume = "584",
    pages = "1--105",
    year = "2015"
}

@article{Gao:2021xsm,
    author = "Gao, Xiang and Karthik, Nikhil and Mukherjee, Swagato and Petreczky, Peter and Syritsyn, Sergey and Zhao, Yong",
    title = "{Pion form factor and charge radius from lattice QCD at the physical point}",
    eprint = "2102.06047",
    archivePrefix = "arXiv",
    primaryClass = "hep-lat",
    doi = "10.1103/PhysRevD.104.114515",
    journal = "Phys. Rev. D",
    volume = "104",
    number = "11",
    pages = "114515",
    year = "2021"
}

@article{Kwee:2007dd,
    author = "Kwee, Herry J. and Lebed, Richard F.",
    title = "{Pion form-factors in holographic QCD}",
    eprint = "0708.4054",
    archivePrefix = "arXiv",
    primaryClass = "hep-ph",
    doi = "10.1088/1126-6708/2008/01/027",
    journal = "JHEP",
    volume = "01",
    pages = "027",
    year = "2008"
}

@article{Chen:2025kqb,
    author = "Chen, Xun and Chen, Yidian and Zhou, Kai",
    title = "{Data-driven Einstein-dilaton model for pure Yang-Mills thermodynamics and glueball spectrum}",
    eprint = "2507.06729",
    archivePrefix = "arXiv",
    primaryClass = "hep-ph",
    doi = "10.1103/wyr3-2kc5",
    journal = "Phys. Rev. D",
    volume = "112",
    number = "12",
    pages = "126025",
    year = "2025"
}

@misc{Cao:2025dkv,
    author = "Cao, Xiong-Hui and Guo, Feng-Kun and Li, Qu-Zhi and Wu, Bo-Wen and Yao, De-Liang",
    title = "{Gravitational form factors of pions, kaons and nucleons from dispersion relations}",
    eprint = "2507.05375",
    archivePrefix = "arXiv",
    primaryClass = "hep-ph",
    doi = "10.1140/epjs/s11734-025-02025-9",
    month = "7",
    year = "2025"
}

@article{Xu:2023izo,
    author = "Xu, Yin-Zhen and Ding, Minghui and Raya, Kh{\'e}pani and Roberts, Craig D. and Rodr{\'\i}guez-Quintero, Jos{\'e} and Schmidt, Sebastian M.",
    title = "{Pion and kaon electromagnetic and gravitational form factors}",
    eprint = "2311.14832",
    archivePrefix = "arXiv",
    primaryClass = "hep-ph",
    reportNumber = "NJU-INP 081/23",
    doi = "10.1140/epjc/s10052-024-12518-x",
    journal = "Eur. Phys. J. C",
    volume = "84",
    number = "2",
    pages = "191",
    year = "2024"
}

@article{Broniowski:2024oyk,
    author = "Broniowski, Wojciech and Ruiz Arriola, Enrique",
    title = "{Gravitational form factors of the pion and meson dominance}",
    eprint = "2405.07815",
    archivePrefix = "arXiv",
    primaryClass = "hep-ph",
    doi = "10.1016/j.physletb.2024.139138",
    journal = "Phys. Lett. B",
    volume = "859",
    pages = "139138",
    year = "2024"
}

@article{Brodsky:2007hb,
    author = "Brodsky, Stanley J. and de Teramond, Guy F.",
    title = "{Light-Front Dynamics and AdS/QCD Correspondence: The Pion Form Factor in the Space- and Time-Like Regions}",
    eprint = "0707.3859",
    archivePrefix = "arXiv",
    primaryClass = "hep-ph",
    reportNumber = "SLAC-PUB-12554",
    doi = "10.1103/PhysRevD.77.056007",
    journal = "Phys. Rev. D",
    volume = "77",
    pages = "056007",
    year = "2008"
}

@article{Brodsky:2008pf,
    author = "Brodsky, Stanley J. and de Teramond, Guy F.",
    title = "{Light-Front Dynamics and AdS/QCD Correspondence: Gravitational Form Factors of Composite Hadrons}",
    eprint = "0804.0452",
    archivePrefix = "arXiv",
    primaryClass = "hep-ph",
    reportNumber = "SLAC-PUB-13192",
    doi = "10.1103/PhysRevD.78.025032",
    journal = "Phys. Rev. D",
    volume = "78",
    pages = "025032",
    year = "2008"
}

@article{Seth:2012nn,
    author = "Seth, Kamal K. and Dobbs, S. and Metreveli, Z. and Tomaradze, A. and Xiao, T. and Bonvicini, G.",
    title = "{Electromagnetic Structure of the Proton, Pion, and Kaon by High-Precision Form Factor Measurements at Large Timelike Momentum Transfers}",
    eprint = "1210.1596",
    archivePrefix = "arXiv",
    primaryClass = "hep-ex",
    doi = "10.1103/PhysRevLett.110.022002",
    journal = "Phys. Rev. Lett.",
    volume = "110",
    number = "2",
    pages = "022002",
    year = "2013"
}

@article{ParticleDataGroup:2022pth,
    author = "Workman, R. L. and others",
    collaboration = "Particle Data Group",
    title = "{Review of Particle Physics}",
    doi = "10.1093/ptep/ptac097",
    journal = "PTEP",
    volume = "2022",
    pages = "083C01",
    year = "2022"
}

@article{Rinaldi:2022dyh,
    author = "Rinaldi, Matteo and Ceccopieri, Federico Alberto and Vento, Vicente",
    title = "{The pion in the graviton soft-wall model: phenomenological applications}",
    eprint = "2204.09974",
    archivePrefix = "arXiv",
    primaryClass = "hep-ph",
    doi = "10.1140/epjc/s10052-022-10538-z",
    journal = "Eur. Phys. J. C",
    volume = "82",
    number = "7",
    pages = "626",
    year = "2022"
}

@article{Li:2021jqb,
    author = "Li, Yang and Vary, James P.",
    title = "{Light-front holography with chiral symmetry breaking}",
    eprint = "2103.09993",
    archivePrefix = "arXiv",
    primaryClass = "hep-ph",
    doi = "10.1016/j.physletb.2021.136860",
    journal = "Phys. Lett. B",
    volume = "825",
    pages = "136860",
    year = "2022"
}

@article{Frezzotti:2008dr,
    author = "Frezzotti, R. and Lubicz, V. and Simula, S.",
    collaboration = "ETM",
    title = "{Electromagnetic form factor of the pion from twisted-mass lattice QCD at N(f) = 2}",
    eprint = "0812.4042",
    archivePrefix = "arXiv",
    primaryClass = "hep-lat",
    reportNumber = "RM3-TH-08-15, ROM2F-2008-24",
    doi = "10.1103/PhysRevD.79.074506",
    journal = "Phys. Rev. D",
    volume = "79",
    pages = "074506",
    year = "2009"
}

@article{QCDSFUKQCD:2006gmg,
    author = {Br{\"o}mmel, D. and others},
    collaboration = "QCDSF/UKQCD",
    title = "{The Pion form-factor from lattice QCD with two dynamical flavours}",
    eprint = "hep-lat/0608021",
    archivePrefix = "arXiv",
    reportNumber = "DESY-06-135, EDINBURGH-2006-17, LTH-710, TUM-T39-06-06",
    doi = "10.1140/epjc/s10052-007-0295-6",
    journal = "Eur. Phys. J. C",
    volume = "51",
    pages = "335--345",
    year = "2007"
}

@article{Brandt:2013dua,
    author = {Brandt, Bastian B. and J{\"u}ttner, Andreas and Wittig, Hartmut},
    title = "{The pion vector form factor from lattice QCD and NNLO chiral perturbation theory}",
    eprint = "1306.2916",
    archivePrefix = "arXiv",
    primaryClass = "hep-lat",
    reportNumber = "HIM-2013-03, MITP-13-029, SHEP",
    doi = "10.1007/JHEP11(2013)034",
    journal = "JHEP",
    volume = "11",
    pages = "034",
    year = "2013"
}

@article{Ding:2024lfj,
    author = "Ding, Heng-Tong and Gao, Xiang and Hanlon, Andrew D. and Mukherjee, Swagato and Petreczky, Peter and Shi, Qi and Syritsyn, Sergey and Zhang, Rui and Zhao, Yong",
    title = "{QCD Predictions for Meson Electromagnetic Form Factors at High Momenta: Testing Factorization in Exclusive Processes}",
    eprint = "2404.04412",
    archivePrefix = "arXiv",
    primaryClass = "hep-lat",
    doi = "10.1103/PhysRevLett.133.181902",
    journal = "Phys. Rev. Lett.",
    volume = "133",
    number = "18",
    pages = "181902",
    year = "2024"
}

@article{NA7:1986vav,
    author = "Amendolia, S. R. and others",
    editor = "Loken, S. C.",
    collaboration = "NA7",
    title = "{A Measurement of the Space - Like Pion Electromagnetic Form-Factor}",
    reportNumber = "CERN-EP-86-34",
    doi = "10.1016/0550-3213(86)90437-2",
    journal = "Nucl. Phys. B",
    volume = "277",
    pages = "168",
    year = "1986"
}

@article{JeffersonLabFpi:2000nlc,
    author = "Volmer, J. and others",
    collaboration = "Jefferson Lab F(pi)",
    title = "{Measurement of the Charged Pion Electromagnetic Form-Factor}",
    eprint = "nucl-ex/0010009",
    archivePrefix = "arXiv",
    reportNumber = "JLAB-PHY-00-16",
    doi = "10.1103/PhysRevLett.86.1713",
    journal = "Phys. Rev. Lett.",
    volume = "86",
    pages = "1713--1716",
    year = "2001"
}

@article{Arnold:1974ik,
    author = "Arnold, C. L. and Roe, B. P. and Sinclair, D.",
    title = "{Measurement of the form-factors in the decay k+ ---{\ensuremath{>}} pi mu nu}",
    doi = "10.1103/PhysRevD.9.1221",
    journal = "Phys. Rev. D",
    volume = "9",
    pages = "1221--1229",
    year = "1974"
}

@article{Hackett:2023nkr,
    author = "Hackett, Daniel C. and Oare, Patrick R. and Pefkou, Dimitra A. and Shanahan, Phiala E.",
    title = "{Gravitational form factors of the pion from lattice QCD}",
    eprint = "2307.11707",
    archivePrefix = "arXiv",
    primaryClass = "hep-lat",
    reportNumber = "MIT-CTP/5585",
    doi = "10.1103/PhysRevD.108.114504",
    journal = "Phys. Rev. D",
    volume = "108",
    number = "11",
    pages = "114504",
    year = "2023"
}

@article{Ananthanarayan:2022wsl,
    author = "Ananthanarayan, B.",
    title = "{Pions: the original Nambu{\textendash}Goldstone bosons: An introduction and precision pion physics}",
    doi = "10.1140/epjs/s11734-022-00443-7",
    journal = "Eur. Phys. J. ST",
    volume = "231",
    number = "2",
    pages = "91--102",
    year = "2022"
}

@article{Gherghetta:2009ac,
    author = "Gherghetta, Tony and Kapusta, Joseph I. and Kelley, Thomas M.",
    title = "{Chiral symmetry breaking in the soft-wall AdS/QCD model}",
    eprint = "0902.1998",
    archivePrefix = "arXiv",
    primaryClass = "hep-ph",
    doi = "10.1103/PhysRevD.79.076003",
    journal = "Phys. Rev. D",
    volume = "79",
    pages = "076003",
    year = "2009"
}

@article{Aliev:2023tqy,
    author = "Aliev, T. M. and Bilmis, S. and Savci, M.",
    title = "{Analysis of the light JP=3- mesons in QCD sum rules}",
    eprint = "2307.00577",
    archivePrefix = "arXiv",
    primaryClass = "hep-ph",
    doi = "10.1103/PhysRevD.108.034020",
    journal = "Phys. Rev. D",
    volume = "108",
    number = "3",
    pages = "034020",
    year = "2023"
}

@article{Brodsky:2006uqa,
    author = "Brodsky, Stanley J. and de Teramond, Guy F.",
    title = "{Hadronic spectra and light-front wavefunctions in holographic QCD}",
    eprint = "hep-ph/0602252",
    archivePrefix = "arXiv",
    reportNumber = "SLAC-PUB-11716",
    doi = "10.1103/PhysRevLett.96.201601",
    journal = "Phys. Rev. Lett.",
    volume = "96",
    pages = "201601",
    year = "2006"
}

@article{deTeramond:2005su,
    author = "de Teramond, Guy F. and Brodsky, Stanley J.",
    title = "{Hadronic spectrum of a holographic dual of QCD}",
    eprint = "hep-th/0501022",
    archivePrefix = "arXiv",
    reportNumber = "SLAC-PUB-10789, NSF-KITP-04-131",
    doi = "10.1103/PhysRevLett.94.201601",
    journal = "Phys. Rev. Lett.",
    volume = "94",
    pages = "201601",
    year = "2005"
}

@article{Bakulev:2004cu,
    author = "Bakulev, A. P. and Passek-Kumericki, K. and Schroers, W. and Stefanis, N. G.",
    title = "{Pion form-factor in QCD: From nonlocal condensates to NLO analytic perturbation theory}",
    eprint = "hep-ph/0405062",
    archivePrefix = "arXiv",
    reportNumber = "IRB-TH-02-04, MIT-CTP-3478, RUB-TPII-02-04",
    doi = "10.1103/PhysRevD.70.033014",
    journal = "Phys. Rev. D",
    volume = "70",
    pages = "033014",
    year = "2004",
    note = "[Erratum: Phys.Rev.D 70, 079906 (2004)]"
}

@article{Chen:2023byr,
    author = "Chen, Long-Bin and Chen, Wen and Feng, Feng and Jia, Yu",
    title = "{Next-to-Next-to-Leading-Order QCD Corrections to Pion Electromagnetic Form Factors}",
    eprint = "2312.17228",
    archivePrefix = "arXiv",
    primaryClass = "hep-ph",
    doi = "10.1103/PhysRevLett.132.201901",
    journal = "Phys. Rev. Lett.",
    volume = "132",
    number = "20",
    pages = "201901",
    year = "2024",
    note = "[Erratum: Phys.Rev.Lett. 134, 229901 (2025)]"
}

@article{Wang:2025irh,
    author = "Wang, Sheng-Quan and Liao, Zuo-Fen and Shen, Jian-Ming and Zhou, Hua and Zhang, Jia-Wei and Yan, Jiang and Wu, Xing-Gang and Di Giustino, Leonardo",
    title = "{Analysis of the pion electromagnetic form factor with next-to-next-to-leading order QCD corrections}",
    eprint = "2507.20479",
    archivePrefix = "arXiv",
    primaryClass = "hep-ph",
    doi = "10.1140/epjc/s10052-025-15174-x",
    journal = "Eur. Phys. J. C",
    volume = "85",
    number = "12",
    pages = "1435",
    year = "2025"
}

@article{Chang:2013nia,
    author = {Chang, L. and Clo{\"e}t, I. C. and Roberts, C. D. and Schmidt, S. M. and Tandy, P. C.},
    title = "{Pion electromagnetic form factor at spacelike momenta}",
    eprint = "1307.0026",
    archivePrefix = "arXiv",
    primaryClass = "nucl-th",
    doi = "10.1103/PhysRevLett.111.141802",
    journal = "Phys. Rev. Lett.",
    volume = "111",
    number = "14",
    pages = "141802",
    year = "2013"
}

@article{Hatta:2025ryj,
    author = "Hatta, Yoshitaka and Schoenleber, Jakob",
    title = "{Sullivan Process near Threshold and the Pion Gravitational Form Factors}",
    eprint = "2502.12061",
    archivePrefix = "arXiv",
    primaryClass = "hep-ph",
    doi = "10.1103/y9fq-y84c",
    journal = "Phys. Rev. Lett.",
    volume = "134",
    number = "25",
    pages = "251901",
    year = "2025"
}

@article{Choi:2025rto,
    author = "Choi, Yongwoo and Son, Hyeon-Dong and Choi, Ho-Meoyng",
    title = "{Gravitational form factors of the pion in the self-consistent light-front quark model}",
    eprint = "2504.14997",
    archivePrefix = "arXiv",
    primaryClass = "hep-ph",
    doi = "10.1103/h7gf-f7p4",
    journal = "Phys. Rev. D",
    volume = "112",
    number = "1",
    pages = "014043",
    year = "2025"
}

@article{Li:2013oda,
    author = "Li, Danning and Huang, Mei",
    title = "{Dynamical holographic QCD model for glueball and light meson spectra}",
    eprint = "1303.6929",
    archivePrefix = "arXiv",
    primaryClass = "hep-ph",
    doi = "10.1007/JHEP11(2013)088",
    journal = "JHEP",
    volume = "11",
    pages = "088",
    year = "2013"
}

@article{MartinContreras:2018nhv,
    author = "Mart{\'\i}n Contreras, Miguel {\'A}ngel and Vega, Alfredo and Cort{\'e}s, Santiago",
    title = "{Light pseudoscalar and axial spectroscopy using AdS/QCD modified soft wall model}",
    eprint = "1811.10731",
    archivePrefix = "arXiv",
    primaryClass = "hep-ph",
    doi = "10.1016/j.cjph.2020.06.018",
    journal = "Chin. J. Phys.",
    volume = "66",
    pages = "715--723",
    year = "2020"
}

@article{Dehghan:2025ncw,
    author = "Dehghan, Zeinab and Almaksusi, F. and Azizi, K.",
    title = "{Mechanical properties of proton using flavor-decomposed gravitational form factors}",
    eprint = "2502.16689",
    archivePrefix = "arXiv",
    primaryClass = "hep-ph",
    doi = "10.1007/JHEP06(2025)025",
    journal = "JHEP",
    volume = "06",
    pages = "025",
    year = "2025"
}
\end{document}